\documentstyle[12pt]{article}
\begin{document}

\title{Time-varying Quark Masses and \\Cosmological Axion Energy}

\author{{\bf S.M. Barr and Bumseok Kyae}\\ Bartol Research Institute\\
University of Delaware\\ Newark, DE 19716}

\date{}
\maketitle

\begin{abstract}

The possibility is examined that the masses of the light quarks $u$ and $d$ 
have varied over the course of the universe's evolution. Such a variation
would have an effect on axion cosmology, and can be the basis of a solution
of the so-called axion energy problem. If this is the case, a scalar
force somewhat weaker than gravity could exist that has a range
of order 1 km. 

\end{abstract}

There is a long history of speculation about fundamental ``constants" of
nature varying over cosmic time scales. It began with a suggestion of 
Dirac that time-varying constants might explain one of the large 
number coincidences in cosmology \cite{dirac}. A motivation for more recent
speculations is that in various kinds of theories the gauge couplings, 
$G_N$, and other parameters can depend on upon the size of extra
space dimensions, or upon the value of dilaton, moduli, or other
fields. Such theories include Kaluza-Klein theories \cite{kaluzaklein}, 
Brans-Dicke \cite{bransdicke} and extended inflation models \cite{extended},
``induced gravity" models \cite{induced}, 
quintessence models \cite{quintessence}, superstring theory and brane
scenarios. From the empirical side, there have been both claims from
time to time of evidence of changing ``constants" and stringent
bounds on such changes. (For a thorough review see \cite{uzan}.)

In order to have ``constants" change over cosmic time scales, there typically
needs to be a field with a very flat potential. Specifically, if such a field
changes by a factor of order unity over the age of the universe, it 
should have a mass of order the Hubble parameter, which today is about
$10^{-42}$ GeV. In this paper, we look at the possibility that the masses of 
one or both of the light quarks $u$ and $d$ is time-varying. 
(Most of our discussion will deal with the case where both $m_u$ and $m_d$
vary with time. However, most of the effects we shall discuss would be the 
same if only one of them did.) We 
assume that they depend on a field $\phi$ that has a very flat potential, and
whose couplings are proportional to $M_{P\ell}^{-1}$.
The flatness of this potential is limited by the very fact that $\phi$ 
couples to $u$ and $d$, since QCD and quark-loop effects generate a mass for 
$\phi$ that is at least of order $\Lambda^2_{QCD}/M_{P \ell} 
\sim 10^{-20}$ GeV. 
This implies that $\phi$ ought to evolve 
rapidly once the temperature of the universe cools to $O(\Lambda_{QCD})$. One
might think that this prevents interesting consequences for physics today.
However, we shall see that such a time variation of $m_u$ and $m_d$ can help
to resolve the well-known ``axion energy problem" \cite{aep}, i.e. the
problem that in typical invisible axion models too much energy
can get trapped in oscillations of the axion field in the early universe,
unless $f_a$ is less than about $10^{12}$ GeV. 
(A variety of other solutions to this problem have been proposed in the
literature \cite{other}.)

The point is that the potential energy of the axion field, which is generated
by QCD instanton effects, vanishes if any of the quark masses vanishes.
Thus, if at the time when the axion potential was ``turning on" (when
$T \sim \Lambda_{QCD}$) the light quark masses were much smaller than they
are today, the axion potential energy would have been correspondingly 
suppressed. It is crucial, of course, that the light quark masses already
had their present values by the time of Big Bang nucleosynthesis, as otherwise
the primordial Helium abundance would come out wrong. On the other hand,
to help resolve the axion energy problem, the light quark masses
must {\it not} already have evolved to their present values before instanton
effects became important at $T \sim \Lambda_{QCD}$. That means that it is 
necessary that the field $\phi$ started to change significantly 
at about the time when 
$T \sim \Lambda_{QCD}$, not much earlier or much later. 
But this implies a mass for $\phi$ of order
$\Lambda_{QCD}^2/M_{P \ell}$, which is just the order that is 
expected to arise from its
coupling to the light quarks, as we have already noted. 

Let us take the Yukawa couplings of the $u$ and $d$ quarks to be of the
form
\begin{equation}
{\cal L}_{Yuk} = Y_u \overline{u_L} u_R H_u (\phi/M_{P \ell}) +
Y_d \overline{d_L} d_R H_d (\phi/M_{P \ell}).
\end{equation}

\noindent
We assume that $\phi$ couples to no other fields of the standard model,
except through $u$ and $d$ quark loops. The phase of $\phi$ is
the axion field, so $\phi$ has a non-zero Peccei-Quinn charge. Let us write
$\phi = \rho(x) e^{i \theta(x)}$. The radial field $\rho$ is assumed to be
sitting now at the minimum $\rho_0$ of its potential. The values of 
$m_u$ and $m_d$ that are observed today we denote by
$m_{u0}$ and $m_{d0}$. For general values of $\rho$, one then has, from 
Eq. (1),
$m_u (\rho) = m_{u0} \rho/\rho_0$ and $m_d (\rho) = m_{d0} \rho/\rho_0$.
The Lagrangian density for $\rho$ and $\theta$ can be written approximately 
as
\begin{equation}
{\cal L}(\rho, \theta) \cong \frac{1}{2} \rho^2(\partial_{\mu} \theta)^2
- \frac{1}{2} m^2_a \rho^2 \theta^2 
+ \frac{1}{2}( \partial_{\mu} \rho)^2 - V(\rho).
\end{equation}

\noindent
The axion mass depends on temperature
as well as on the value of the radial field $\rho$, so we will write
$m^2_a (\rho, T) = c_a (T) m^2_a (\rho, 0)$, where $c_a (T) \cong 1$ for $T
\ll \Lambda_{QCD}$ and $c_a (T)$ falls off rapidly for
$T \gg \Lambda_{QCD}$ \cite{gpy}. 
The axion mass depends on $\rho$ both because of
the fact that $a(x) = \rho(x) \theta(x)$ and because of the $\rho$ dependence
of $m_u$ and $m_d$. The axion mass can be obtained from the term in the chiral
Lagrangian that also gives the neutral pion mass, namely
\begin{equation}
-C[m_u(\rho) e^{i(\theta + \pi^0/f_{\pi})} +
m_d(\rho) e^{i(\theta - \pi^0/f_{\pi})}] + H.c. 
\end{equation}

\noindent
This leads to the zero-temperature potential
\begin{equation}
V(\rho) = -2 C (m_u(\rho) + m_d(\rho)) 
+ C( m_u(\rho)(\theta +  \pi^0/f_{\pi})^2 + 
m_d(\rho) (\theta - \pi^0/f_{\pi})^2) + ...
\end{equation}

\noindent
After diagonalizing the mass$^2$ matrix of $\theta$ and $\pi^0/f_{\pi}$,
one has $m^2_{\pi}(\rho) = 2 C(m_u(\rho) + m_d(\rho))/f^2_{\pi}$, and
$m^2_a (\rho, 0) = 
8 C \rho^{-2} m_u (\rho) m_d (\rho)/[m_u (\rho) + m_d (\rho)]$,
giving
\begin{equation}
m^2_a = 4 \frac{m^2_{\pi} f^2_{\pi}}{\rho^2} 
\frac{m_u(\rho) m_d(\rho)}{(m_u(\rho) + m_d(\rho))(m_{u0} + m_{d0})},
\end{equation}

\noindent
which goes as $\rho^{-1}$ for small $\rho$, and 
\begin{equation}
V_{QCD}(\rho) = - m^2_{\pi} f^2_{\pi} \left(
\frac{m_u (\rho) + m_d (\rho)}{m_{u0} + m_{d0}} \right)'
\end{equation}

\noindent
which goes as $\rho^{+1}$ for small $\rho$.
Not only the axion mass but also the potential for $\rho$ generated by chiral 
symmetry breaking depends on temperature. So we will write
$V_{QCD}(\rho, T) = - c_{\rho} (T) V_{QCD} (\rho, 0)$,
where $c_{\rho}(T)$ equals 1 below $\Lambda_{QCD}$ and falls off rapidly
above it. There are also contributions to $V(\rho)$ 
coming from $u$ and $d$ quark loops. These go as 
\begin{equation}
V_{1-loop} \sim -\frac{1}{16 \pi^2} (|Y_u|^2 |H_u|^2 + |Y_d|^2 |H_d|^2)
\rho^2 \Lambda^2/M^2_{P \ell} = - \frac{1}{16 \pi^2} (|m_u(\rho)|^2 
+ |m_d(\rho)|^2) \Lambda^2,
\end{equation}

\noindent
where $\Lambda$ is a cutoff of order the weak scale. (The same loops
contribute to the (mass)$^2$ 
of the Higgs doublets, so that whatever solves the gauge 
hierarchy problem should cut off the loops at the weak scale.) The expressions
given in Eqs. (6) and (7) are roughly comparable for $\rho \sim \rho_0
\sim M_{P \ell}$ and are the leading contributions coming from the coupling of 
$\rho$ to the $u$ and $d$ quarks. They both give a mass scale for
$\rho$ that is
roughly of order $\Lambda_{QCD}^2/M_{P \ell}$. As noted above, this is the kind
of mass that is required for the mechanism we are discussing for suppressing
the cosmological axion energy density. Thus it is important that there
not be much larger contributions to $m^2_{\rho}$ coming from other sources, 
such as supersymmetry breaking. 

The contributions to $V(\rho)$ in Eqs. (6) and (7) are both unbounded below.
It is therefore necessary that there be other contributions to $V(\rho)$.
We assume that there are corrections to $V(\rho)$ coming from Planck-scale
physics that cause it to have a minimum at $\rho_0 \sim M_{P \ell}$. In 
particular, suppose that there are corrections to the terms in Eq. (1) 
of order $|\phi|^2/M^2_{P \ell}$ as follows: $Y_u \overline{u_L} u_R H_u \phi
(1 - a_u |\phi|^2/M^2_{P \ell}) + Y_d \overline{u_L} u_R H_d \phi
(1 - a_d |\phi|^2/M^2_{P \ell})$. In this case, if $a_{u,d}$ are positive,
$|m_u(\rho)|$ and $|m_d(\rho)|$ have maxima and therefore the terms in Eqs. (6) and (7), which go respectively as $-|m_{u,d}(\rho)|^1$ and 
$-|m_{u,d}(\rho)|^2$, have minima. To simplify the discussion, let us assume 
that $a_u = a_d \equiv a$ and that $V(\rho) \sim |m_{u,d}|^n$ near its minimum.
Then we can write $m_q(\rho) \cong \frac{3}{2} m_{q0} (\rho/\rho_0)
(1 - \frac{1}{3} \rho^2/\rho_0^2)$, $q = u,d$, as this leads to a maximum for 
$|m_q|$ (and thus a minimum for $V(\rho)$) at $\rho_0$, and gives 
$m_q(\rho_0) = m_{q0}$. Then we may reexpress the axion mass as
\begin{equation}
m_a^2(\rho, 0) \cong 6 \frac{(1 - \rho^2 /3 \rho_0^2)}{\rho_0 \rho}
\frac{m_{\pi}^2 f_{\pi}^2 m_{u0} m_{d0}}{(m_{u0} + m_{d0})^2}.
\end{equation}

\noindent
For $\rho \ll \rho_0$, where $V_{QCD}$ dominates over $V_{1-loop}$, one has
\begin{equation}
V(\rho, 0) \cong - \frac{3}{2} m^2_{\pi} f^2_{\pi} \left( \frac{\rho}{\rho_0} 
\right).
\end{equation}
We are now in a position to disuss the suppression of cosmological
axion energy. First, let us see in rough terms what happens. The QCD instanton
effects begin to be important when $T \sim \Lambda_{QCD}$. At that point,
the Hubble parameter $H^2 \sim \Lambda_{QCD}^4/M^2_{P \ell}$, and from Eq. (8),
$m^2_a( \rho, T) = c_a (T) m^2_a (\rho, 0) \sim c_a (T) 
\frac{m^2_{\pi} f^2_{\pi}}{\rho_0 \rho} \sim c_a (T) 
(\Lambda_{QCD}^4/M^2_{P \ell})
(\rho_0/\rho)$. Thus, the axion oscillations commence when
$c_a (T) \sim \rho_i/\rho_0 \ll 1$, $\rho_i$ being the initial value of $\rho$.
The energy density in the axion oscillations at that point is
$\rho_i^2 m_a^2(\rho_i, T_i) \sim (\rho_i^2/\rho_0^2) \Lambda_{QCD}^4 \sim
(\rho_i^2/M_{P \ell}^2) \Lambda_{QCD}^4$. In other words, it is suppressed by
a factor $(\rho_i/M_{P \ell})^2$ compared to what it would be in a conventional invisible axion model having $f_a \sim M_{P \ell}$. 
Another thing that happens when $T 
\sim \Lambda_{QCD}$ is that the radial field $\rho$ begins to roll due to the
potential energy shown in Eq. (9). As $\rho$ changes, the axion mass changes,
both because $a = \rho \theta$ and because of the $\rho$ dependence
of $m_u$ and $m_d$. The {\it number} of axions does not change significantly 
during this process, however, since the axion mass is varying adiabatically.
Consequently, the number density of axions falls as $R^{-3}$, where $R$ is the
cosmic scale factor. The mass of the axion ``initially" (i.e. when axion
oscillations start) is of order $H \sim \Lambda_{QCD}^2/M_{P \ell}$, 
whereas by the
time $\rho$ has settled at its minimum the axion mass has reached its present
value of $(m_{\pi} f_{\pi}/\rho_0)$, which is also of order 
$\Lambda_{QCD}^2/M_{P \ell}$. Thus, $(m_a)_f/(m_a)_i \sim 1$.
Therefore, over the whole period when $\rho$ is
evolving, the axion energy density changes by about the same factor as the 
axion number density, namely by $(R_i/R_f)^3$. It follows that the initial
suppression of the axion energy density by $(\rho_i/M_{P \ell})^2$ (compared
to conventional axion models with $f_a \sim M_{P \ell}$) will also be 
approximately
the suppression of the final axion energy density. Since a suppression factor
of about $10^{-8}$ is required, it is sufficient to have $\rho_i \sim 10^{-4}
M_{P \ell}$, which is of order the unification scale.

Now we will be more detailed. The equation for the rolling of the radial field 
is
\begin{equation}
\ddot{\rho} + 3 H \dot{\rho} = - V'(\rho).
\end{equation}

\noindent
The condition for when $\rho$ starts to roll is that $\dot{\rho} \sim
H \rho$ and $\ddot{\rho} \sim H^2 \rho$. This gives
\begin{equation}
4 H^2 \rho_i \sim c_{\rho}(T) m_{\pi}^2 f_{\pi}^2/\rho_0,
\end{equation}

\noindent
where the initial value of $\rho$ is $\rho_i$. Calling the temperature 
at that time $T_{\rho}$, defining $T_{QCD} \equiv \sqrt{m_{\pi} f_{\pi}}$,
and using $H^2 = \frac{8 \pi}{3} \left( \frac{\pi^2}{30} g T^4 \right)
/M^2_{P \ell}$, 
where g is the effective number of polarizations of radiation, we have 
\begin{equation}
c_{\rho}(T_{\rho})^{-1} (T_{\rho}/T_{QCD})^4 \cong \frac{45}{16 \pi^3 g} 
\left( \frac{M_{P \ell}}{\rho_0}
\right)^2 \left( \frac{\rho_0}{\rho_i} \right).
\end{equation}

\noindent
The axion field starts oscillating at a temperature $T_a$ given by
\begin{equation}
m^2_a(\rho_i, T_a) = c_a (T_a) m^2_a(\rho_i, 0) = H^2 = \frac{8 \pi}{3}
\left( \frac{\pi^2}{30} g T_a^4 \right)/M_{P \ell}^2 ,
\end{equation}

\noindent 
or
\begin{equation}
c_a (T_a)^{-1} (T_a/T_{QCD})^4 \approx \frac{135}{2 \pi^3 g} \left( 
\frac{M_{P \ell}}{\rho_0} \right)^2
\left( \frac{\rho_0}{\rho_i} \right) \frac{m_{u0} m_{d0}}{(m_{u0} + m_{d0})^2}.
\end{equation}

\noindent
Comparing Eqs. (12) and (14), we see that the axion oscillations begin
at about the same time that the radial field starts to roll, a time when
the temperature is slightly higher than $T_{QCD}$. (We will find that
$\rho_0/\rho_i \sim 10^3$.) Initially, that is at $T = T_a$, the energy 
in axion oscllations is
$(\rho_{axion})_i \approx \frac{1}{2} c_a (T_a) m_a^2( \rho_i, 0) \rho_i^2 
\theta_i^2$, where $\theta$ is measured from the minimum of the potential.
By Eq. (13) this is
\begin{equation}
(\rho_{axion})_i \approx \frac{2 \pi^3}{45} g T^4_a \left( 
\frac{\rho_i}{M_{P \ell}}
\right)^2 \theta_i^2.
\end{equation}

\noindent
The baryon number density at that time was 
\begin{equation}
(n_B)_i = \left( \frac{n_B}{s} \right)_i \left( g \frac{2 \pi^2}{45} 
T_a^3 \right).
\end{equation}

\noindent
Assuming adiabatic evolution since, one has $(n_B/s)_i = (n_B/s)_{now}
= (n_B/n_{\gamma})_{now}/7.04  \equiv \eta_B/7.04$, and from Eqs. (15) and 
(16),
\begin{equation}
\left( \frac{\rho_{axion}}{\rho_B} \right)_{now} \approx
\frac{(m_a)_f}{(m_a)_i} \frac{T_a}{m_p} \frac{7.04 \pi}{\eta_B} 
\left( \frac{\rho_i}{M_{P \ell}} \right)^2 \theta_i^2,
\end{equation}

\noindent
where $m_p$ is the proton mass. 
However, we know from Eq. (5) that $(m_a^2)_f = 
4 \frac{m^2_{\pi} f^2_{\pi}}{\rho_0^2} 
\frac{m_{u0} m_{d0}}{(m_{u0} + m_{d0})^2}$. Dividing by Eq. (13):
\begin{equation}
\frac{(m_a)_f}{(m_a)_i} \approx \left( \frac{45}{\pi^3 g} \right)^{1/2}
\left( \frac{T_{QCD}}{T_a} \right)^2 
\frac{\sqrt{m_{u0} m_{d0}}}{m_{u0} + m_{d0}} \left( \frac{M_{P \ell}}{\rho_0}
\right).
\end{equation}

\noindent
From Eqs. (17), (18), and (14), one has
\begin{equation}
\left( \frac{\rho_{axion}}{\rho_B} \right)_{now} \approx
\left[ \sqrt{\frac{45}{\pi g}} \frac{\sqrt{m_{u0} m_{d0}}}{m_{u0} + m_{d0}}
\frac{T_{QCD}}{m_p} \frac{7.04}{\eta_B} \right] \left( \frac{T_{QCD}}{T_a}
\right) \left( \frac{\rho_i}{\rho_0} \right)^2 \left( 
\frac{\rho_0}{M_{P \ell}} \right) \theta_i^2.
\end{equation}

\noindent
Using $c_a (T) \sim (T/T_{QCD})^{-5}$ 
\cite{gpy}, one then has
\begin{equation}
\left( \frac{\rho_{axion}}{\rho_B}\right)_{now} \approx
10^9 \left( \frac{\rho_i}{\rho_0} \right)^{19/9} 
\left( \frac{\rho_0}{M_{P \ell}} \right)^{11/9} \theta_i^2.
\end{equation}

\noindent
We require that at the present time $\Omega_{axion}/\Omega_B$ be less
than about 10. This can be achieved with $\rho_i \approx 10^{-4} \rho_0$ to
$10^{-3} \rho_0$, which is what we found from our earlier, cruder discussion.

We have been assuming that both $m_u$ and $m_d$ depend on $\rho$. It is also
possible to imagine that $\phi$ couples only to $u$ or only to $d$. The results
would be qualitatively the same. The axion mass vanishes if any one
quark mass goes to zero. And the QCD chiral symmetry breaking effects would
still lead to a linear potential for $\rho$ given by Eq. (6) with either
$m_d (\rho)$ replaced by $m_{d0}$ or $m_u (\rho)$ replaced by $m_{u0}$. 

A potential problem for the scenario of a rolling $\rho$ field
is the energy that may get trapped
in the oscillations of that field. For a realistic cosmology
it is necessary that the $\rho$ particles be able to decay to other
massless or nearly massless particles, whose energy can be red-shifted away.
There is nothing in principle to prevent this.  

Suppose that the decay rate of $\rho$ excitations into massless
particles is given by $\Gamma_{\rho} = \alpha m_{\rho}$, where $\alpha$ is
the appropriate combination of dimensionless couplings and powers of $4\pi$. 
This decay must happen before nucleosynthesis; so that $\alpha$ must 
be greater than about $10^{-4}$.
Then, between the time $\rho$ oscillations begin and the time they are damped 
out by decays, the universe expands by a factor $\approx 
(\alpha)^{-1/2}$, assuming radiation domination. The density of
energy in the $\rho$ field when it begins to
oscillate is of order $|V(\rho_0)|$. The density of radiation at
that time is $\rho_{rad} = \frac{3 M_{P \ell}^2}{8 \pi} 
m^2_{\rho} \sim \frac{3 M_{P \ell}^2}{8 \pi} |V(\rho)|/\rho^2_0$. 
Thus, initially, $\Omega_{\rho}
\sim (\rho_0/M_{P \ell})^2$, and consequently after
the $\rho$ oscillations decay to massless
modes the fraction of energy in those modes is
$\Omega_{\rho dec} \sim (\alpha)^{-1/2} (\rho_0/M_{P \ell})^2$. 
In order for 
this not to be in conflict with bounds from Big Bang nucleosynthesis
on massless degrees of freedom, one must therefore assume that $\rho_0$
is less than about $\alpha^{1/4} M_{P \ell}/10$, which is between about
$10^{-1} M_{P \ell}$ and $10^{-2} M_{P \ell}$. 

A crucial question is whether the radial field $\rho$ mediates a long-range
force that would violate the equivalence principle. The range of such a
force would be $m_{\rho}^{-1} = [V^{\prime \prime} (\rho)]^{-1/2} \sim
\rho_0/\sqrt{V(\rho_0)} \sim \frac{4 \pi \rho_0}{m_{u,d} \Lambda_{Weak}}
\sim 1$ km. The couplings of $\rho$ to the $u$ and $d$ quarks are suppressed 
by $1/\rho_0$, but since this is larger than $1/M_{P\ell}$ there is obviously
a danger of violating current limits on long-range scalar forces.

This danger is neatly avoided in an interesting class of cases, namely where
all the light quarks whose masses depend substantially on $\rho$ depend
on $\rho$ {\it in the same way}. This would trivially be the case if
$\phi$ only coupled to $u$ or only to $d$. If $\phi$ coupled to both
$u$ and $d$, it would mean that $m_u(\rho)/m_{u0} = m_d(\rho)/m_{d0}
= f(\rho)/f(\rho_0)$. (In the example studied above, this would imply that
the constants $a_u$ and $a_d$ defined before Eq. (8) are equal.) Since 
$V(\rho)$ depends monotonically on the light quark masses (see Eqs. (6)
and (7)), and these in turn are proportional to $f(\rho)$, it is evident
that extremizing $V(\rho)$ with respect to $\rho$ also extremizes
$f(\rho)$, and thus extremizes the light quark masses themselves.
That means that if we then expand
$m_u (\rho) \overline{u} u + m_d (\rho) \overline{d} d$ about the minimum
$\rho = \rho_0$, calling the excitations of the radial field $\tilde{\rho}$,
one finds that $\overline{u} u$ and $\overline{d} d$ couple to $\tilde{\rho}^2$
but not $\tilde{\rho}$. In other words there is no fifth force.
In other words, there is a dangerous fifth force 
only if both $m_u$ and $m_d$
depend significantly on $\rho$ and depend on it 
in different ways, or if $V(\rho)$ does not depend monotonically on the quark 
masses. For, in either of these cases, minimizing $V(\rho)$ does not 
extremize the light
quark masses. If there is a residual fifth force, its couplings to nucleons
would be through the gluon field, and thus of equal strength for protons
and neutrons \cite{svz}. 

In conclusion, we have explored the idea that the light quarks may
couple to a field with a very flat potential and may as a consequence
have varied over the history of the universe. We have seen that this opens up
certain interesting possibilities. These include a relaxing of the cosmological
bound on the axion decay constant, and residual fifth forces. The question 
arises, as in many scenarios with parameters varying over cosmic times scales,
why the potential of the slowly varying field is so flat. There is also the
question what sets the initial value of this field. To relax
the cosmological bound on $f_a$ it must start at a value much smaller than
its value at the minimum of its potential.

While this paper was being written, we became aware of a recent paper by
Berkooz, Nir, and Volanski \cite{bnv}, which also considers the 
idea of time-varying
Yukawa couplings of quarks. Their idea was that the quark masses could have 
been larger at the time of the electroweak phase transition in order to
enhance electroweak baryogenesis coming from the CKM phase. The quark masses
in their scenario are proportional to some singlet field, analogous to
the field we call $\phi$. This field gets shifted at the time of the 
electroweak phase transition. This is quite similar to the situation 
we considered, except that we have the shift taking place due to the QCD
chiral phase transition and
suppose the quarks to have been lighter at early times rather than heavier.

\end{document}